\documentclass[twocolumn,showpacs,preprintnumbers,amsmath,amssymb,aps,prl]{revtex4}
\usepackage{graphicx}

\begin{document}

\title{Cooperative Behavior and Pattern Formation in Mixtures of  
Driven and Non-Driven Colloidal Assemblies}  
\author{C. Reichhardt and C.J. Olson Reichhardt} 
\affiliation{
Center for Nonlinear Studies and Theoretical Division,~Los 
Alamos National Laboratory, Los Alamos, New Mexico 87545}

\date{\today}

\begin{abstract}
We simulate a disordered assembly of 
particles interacting through a repulsive
Yukawa potential with a small fraction of the particles 
coupled to an external drive. 
Distortions in the arrangement of the 
nondriven particles produce a dynamically induced effective
attraction between the driven particles,
giving rise to intermittent one-dimensional stringlike structures. 
The velocity of a moving string increases with the number of 
driven particles in the string.     
We identify the average stable string length as a function of 
driving force, background particle density, and particle charge. 
This model represents a new type of collective transport system 
composed of 
interacting particles moving through deformable disorder. 
\end{abstract}
\maketitle
\vskip2pc

The collective transport of interacting particles
driven through disordered backgrounds has been studied extensively in  
systems such as moving 
vortex lattices in the presence of random pinning \cite{Dominguez},
driven charge density waves \cite{Grunner}, and  sliding friction \cite{Braun}.
A variety 
of nonequilibrium behaviors occur in these systems, including
fractal flow patterns, avalanches,   
nonlinear velocity-force responses, and dynamic reordering transitions.  
The collective effects arise due to
the particle-particle interactions, 
while the disorder in the
substrate is quenched and does not change with time. 
It is also possible to have substrates which can
distort and respond to the motion of the driven particles,
such as in the case of 
a single colloidal particle 
driven through a disordered background of other particles
which are not coupled to the external drive  \cite{Weeks,Reichhardt}. 
Both simulations \cite{Reichhardt} and experiments \cite{Weeks}
show that such a system can exhibit a nonzero threshold force
for motion as well as 
a nonlinear velocity vs applied force response. The driven 
particle moves in   an
intermittent manner 
and causes local rearrangements and distortions in the surrounding
bath of colloids.
An open question is what happens if there are multiple particles 
driven through a background of nondriven particles, rather than only
a single driven particle.
In particular, it is possible that the ability of the 
surrounding media to react to the motion
of driven particles could induce 
an emergent effective driven particle-particle interaction,
leading to new types 
of moving structures or to pattern formation that is not observed in 
driven systems with fixed or quenched background disorder.    

Recent studies have focused on a related system 
in which two sets of identical repulsive
particles are driven in opposite directions
\cite{BeateVicsek,Lowen,Helbing,VanBlarden}. 
Here, a laning phenomenon occurs, and
the species segregate into multiple streams flowing past one another. 
The oppositely driven species model has been studied both in 
terms of pedestrian dynamics, where the particles 
represent people moving in opposite directions 
\cite{Helbing}, 
as well as in the context of colloidal or charged Yukawa particles in the
presence of a driving field \cite{Lowen}.
The colloidal systems are particularly attractive 
for further study since 
binary mixtures of oppositely charged colloids which can
undergo laning transitions in the presence of a 
driving field have been produced recently in experiment
\cite{VanBlarden}. This type of
system opens a wealth of new experimental possibilities. 

To our knowledge, the case of a small fraction of particles driven 
through a responsive but undriven background 
has not been studied previously. 
In this work, we show that when multiple particles move
through a background of nondriven particles, an effective emergent attraction
arises between the driven particles despite the fact that all of the
pairwise interactions between the particles are repulsive.
We find that intermittent moving strings form
which have an average stable length that is a function of the
driving force, system density, and 
particle charge.
Strings that are longer than average are very short lived.
The string velocity increases monotonically with the string length.
    
We consider a two-dimensional system with periodic boundary conditions
in the $x$ and $y$ directions
containing $N_c$ particles interacting via a screened Coulomb
or Yukawa interaction. 
Only 
a small number $N_{D}$ of the particles couple to an external drive.
The particles move in an overdamped background, and we neglect 
hydrodynamic effects, which is reasonable for the low volume fraction
limit considered here.  
We use molecular dynamics (MD) to 
numerically integrate the overdamped equations of motion
for the particles, given for particle $i$ by 
$d {\bf r}_{i}/dt = {\bf F}_{i}^{cc}+{\bf F}_{i}^{D} + {\bf F}^{T}_{i}.$
The colloid-colloid interaction  
force term 
${\bf F}_{i}^{cc} = -q_{i}\sum_{j \neq i}^{N_{c}}\nabla_i V(r_{ij})$ 
contains the Yukawa interaction potential 
$V(r_{ij}) = (q_{j}/|{\bf r}_{i} - {\bf r}_{j}|)
\exp(-\kappa|{\bf r}_{i} - {\bf r}_{j}|)$, where
${\bf r}_{i(j)}$ is the position of
particle $i(j)$,
$q_{i(j)}$ is the charge of colloid $i(j)$, and all colloids have
the same sign of charge so that all interaction forces are repulsive.
The screening length $1/\kappa$ is 
set to $1/2$ in all our simulations. 
The background colloids are composed of a 50:50 mixture of two
charges with $q_{1}/q_{2} = 1/2$, which produces a noncrystalline
background arrangement.
The average background charge $q=(q_1+q_2)/2$.
For the non-driven particles,
the constant drive force term ${\bf F}_{D}^{i} = 0$.  
The colloids that couple to the external field have charge $q_{D}$ 
and experience a constant drive ${\bf F}_{D}=F_D{\bf \hat{x}}$ 
applied after the system is equilibrated by simulated annealing. 
The thermal force ${\bf F}_i^T$ has the properties
$\langle {\bf F}_i^T  \rangle=0$ and
$\langle {\bf F}_i^T(t){\bf F}_j^{T}(t^\prime) \rangle=2\pi\eta k_BT\delta_{ij}
\delta(t-t^\prime)$.
For a system of length $L$ the 
colloid density is $n_{c} = N_{c}/L$. We consider both the 
constant $n_c$ case,
where $L$ and $N_c$ are increased simultaneously,
as well as the case 
where $n_c$ is increased by fixing $L$ and increasing $N_{c}$.

\begin{figure}
\includegraphics[width=3.3in]{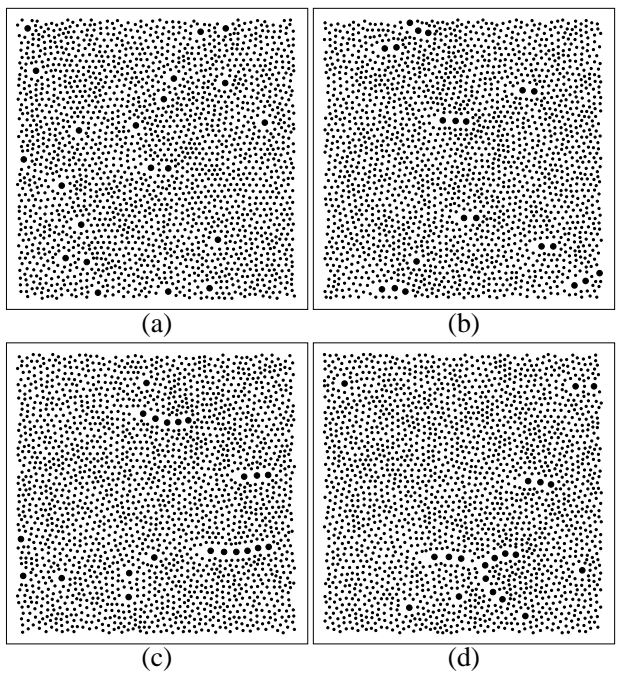}
\vspace{-0.2in}
\caption{
Images of the system composed of a disordered 
array of background colloids (small circles) and
particles that couple to
the external drive (large circles) 
for $q_{D}/q = 6.67$, $n_c = 0.94$, $F_{D} = 3.0$, and
$N_{D}/N_c = 0.01$. (a) Initial configuration; 
(b) the same system after $10^5$ MD steps; (c) a typical steady state 
snapshot after $10^6$ MD steps; (d) snapshot of a string breaking
into shorter strings.          
The driving force is from left to right.
\vspace{-0.2in}
}
\end{figure}

In Fig.~1 we show four snapshots which highlight the 
dynamically induced attraction of the driven particles, indicated as
large black dots, 
and the formation of the string structures. Here the density $n_c = 0.94$,
$q_{D}/q = 6.67$, and $F_{D} = 3.0$. 
Fig.~1(a) shows the initial state in which the        
driven particles are well separated from each other. In Fig.~1(b), 
after $10^5$ MD time steps, 
the driven particles begin
to form string like structures of length $N_s=2$ to 3 colloids
which are aligned in the direction of
the driving force (left to right in the figure). 
In this nonequilibrium system, despite 
the fact that all of the pair-wise interactions
are repulsive, an effective attraction 
emerges between the driven particles.
What is difficult to convey through still images
is the fact that as $N_s$, the number of particles in a string, increases, 
the average string velocity $\langle V_x^s\rangle$ increases 
so that strings  move considerably
faster than isolated driven particles.  
In Fig.~1(c) after $10^6$ MD time steps, 
the system has reached a nonequilibrium steady state. 
Here the average stable length of a string $\langle N_s\rangle=5$ to 6.
Strings with $N_s>6$ form occasionally but
generally have a very short lifetime.
In Fig.~1(c), strings with $N_s=6$, $5$, $3$, and $1$ are present. 
All of the strings have
an {\it intermittent} character, in that driven particles in the string can 
become detached from the string, and new driven particles can join the string.
Typically strings
add or shed one driven particle at a time; however, 
long strings with $N_s=7$ or 8
tend to break up into separate
strings with $N_s=3$, $4$, or $5$, as illustrated in Fig.~1(d).

\begin{figure}
\includegraphics[width=3.5in]{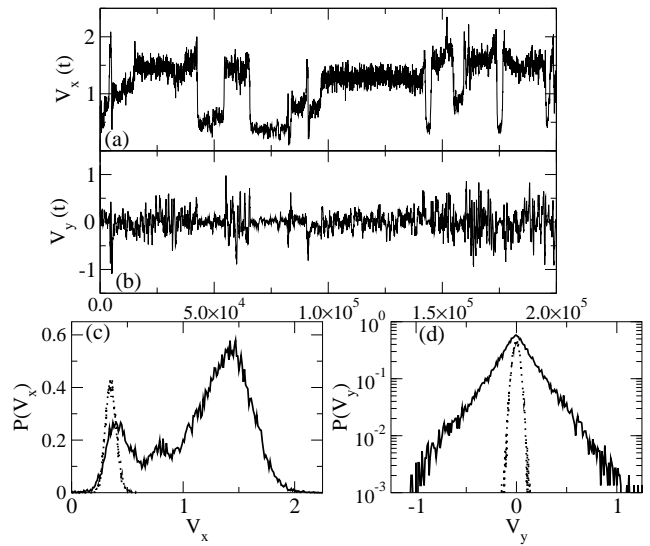}
\vspace{-0.2in}
\caption{
(a) Time series $V_{x}(t)$ for a single driven colloid 
from the system in Fig.~1(a) with $N_D=20$. 
(b) The corresponding time series for $V_{y}$. 
(c) Solid line: The normalized histogram of $V_x$ from the complete time
series for the system in (a).  Dotted line: Histogram of $V_{x}$ for 
the case $N_D=1.$
(d) Solid line: The normalized histogram of $V_{y}$ 
for the total time series from the system in (b). Dotted line:
Histogram of $V_{y}$ for the case $N_D=1$.
\vspace{-0.2in}
}
\end{figure}
               
We note that if $N_D/N_c$ is increased while all other
parameters are held fixed, 
then $\langle N_s\rangle$ remains unchanged at 
$\langle N_s\rangle=5$ to 6.
We find the formation of stable strings over an extensive range
of parameters, including all $q_{D}/q \geq 1.0$ (where we have checked up 
to $q_{D}/q = 40$),
for $n_c > 0.17$, 
and for arbitrarily large $F_{D}$. The 
value of $\langle N_s\rangle$
depends on the specific choice of parameters; however, 
the phenomenon of string formation is very robust
provided that the driven particles
can distort the arrangement of the background particles.
In Fig.~1, $N_D/N\approx 0.01$, 
so that the initial average distance between driven particles is many 
times larger than the pair-interaction length.  Thus, the attractive force 
between driven particles is 
mediated by the distortion field produced in the
background particles.  
If the fraction is increased as much as an order of magnitude up to
$N_D/N_c=0.1$, the preferred length of $\langle N_s\rangle=5$ to 6
persists. We note that if $N_D/N_c$
is too high, finite size effects due to the boundary conditions 
become important. 

In order to illustrate 
the intermittent behavior and the velocity increase as the number of
colloids in a string increases, in Fig.~2(a) we plot a typical 
time series of the velocity in the direction of drive 
$V_x$ for one of the driven 
colloids from the system in Fig.~1(a). Here $V_{x}(t)$ shows a series of 
well defined increasing or decreasing jumps,  
with a roughly constant velocity
between the jumps.
The lower value that $V_{x}$ takes is around $0.39$ 
corresponding to the average velocity
of a driven particle that is not in a string. 
There are several places where $V_{x} = 1.45$ which 
corresponds to the average velocity
of a string with $N_s=5$. Additionally, there are some plateaus
near a value of $V_{x} = 0.7$ which corresponds to $N_s=3$.
Fig.~2(a) also shows that for short times,
$V_x>1.5$, which 
corresponds to strings with $N_s>6$.
This figure shows
that the driven colloid is exiting and joining strings of different lengths.

\begin{figure}
\includegraphics[width=3.5in]{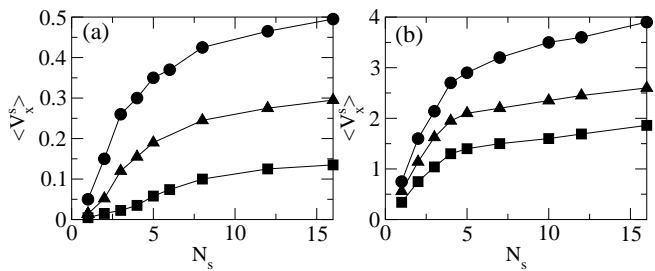}
\vspace{-0.2in}
\caption{$\langle V_{x}^s\rangle$ vs the number of particles in the string 
$N_s$. 
(a) The same system as in Fig.~1(a) 
with $q_D/q=6.67$ for $F_{D} = 3.0$ (squares), 
$4.0$ (triangles), and $5.0$ (circles).
(b) The same system as in (a) for $q_{D}/q = 1.33$ and 
$F_{D} = 0.5$ (squares), $0.75$ (triangles), and $1.0$ (circles).        
\vspace{-0.2in}
}
\end{figure}

Fig.~2(a) is only a portion of a much larger time series
which covers $10^7$ MD steps. 
In Fig.~2(c) we plot the histogram of 
this entire time series along with
the histogram of $V_{x}$ (dotted line) for 
a case where
there is only one driven particle in the system so that strings
do not form.
The $N_D=1$ curve shows a single peak 
in $P(V_x)$ near $V_{x} = 0.35$, 
which falls near but slightly below the first peak in $P(V_x)$ for
the multiple driven particle system.  Thus, the lowest value
of $V_x$ in Fig.~2(a) corresponds to time periods when the driven colloid is
moving individually and is not part of a string.
Additionally, there is a small peak in $P(V_x)$
for the $N_D=20$ system near $V_x=0.8$, 
which corresponds to the formation of strings with $N_s=2$.
There is a large peak in $P(V_x)$ around $V_x=1.45$ 
corresponding to $N_s=5$ and 6.
For $V_{x} > 1.5$ 
there is a rapid decrease in $P(V_x)$. 
This result shows that there is a preferred string
length of $\langle N_s\rangle \approx 5$ for the $N_D=20$ system. 
The average velocity of a driven particle
in the $N_D=20$ case is 
much larger than the average velocity in the $N_D=1$ case, even through
the fraction of driven particles is only $N_D/N_c=0.01$. 

Driven particles traveling in
strings show a more pronounced transverse wandering 
motion than individual driven particles.
In Fig.~2(b), we plot the time series for the velocity transverse to the
driving force, $V_{y}$, for the same particle as in Fig.~2(a). 
Here $\langle V_{y}\rangle = 0$, 
and there are numerous spikelike events 
in the positive and negative directions. In 
Fig.~2(d) we plot the histogram of the entire time series of $V_{y}$ 
(solid line) along with the same histogram 
for a system containing only a single driven particle
(dashed line). The $N_D=1$ curve
shows a well defined Gaussian distribution 
of $P(V_y)$ centered at zero. For $N_D=20$,
the spread in $V_y$ is much larger and the fluctuations 
are non-Gaussian, as indicated by the presence of large tails 
in the distribution. 
There are an excess of events at larger values of $|V_{y}|$ 
which result from the fact that the particles in a chain move 
in a correlated fashion. The non-Gaussian statistics also 
indicate that the strings have transverse superdiffusive behavior.

\begin{figure}
\includegraphics[width=3.5in]{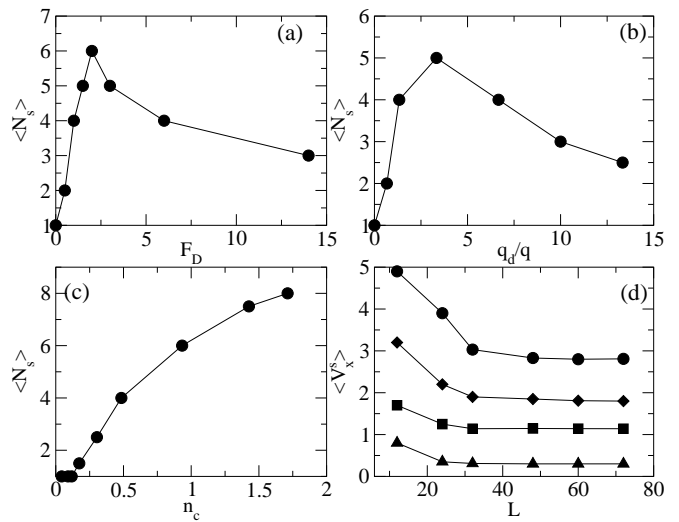}
\vspace{-0.2in}
\caption{
(a) The most probable number of particles in a string 
$\langle N_s\rangle$ vs $F_{D}$ for the system in Fig~1.
(b) $\langle N_s\rangle$ vs $q_{D}$ for the same system as in (a) for 
$F_{D} = 1.0$.     
(c) $\langle N_s\rangle$ vs $n_c$ for $q_{D} = 6.67$ and $F_{D} = 2.0$. 
(d) $\langle V_{x}^s\rangle$ vs system size $L$ for the system in Fig.~1 for 
$F_D=$ 1.5 (triangles), 3.0 (squares), 4.5 (diamonds), and 6.0 (circles).
\vspace{-0.2in}
}
\end{figure}

We next examine the average velocity of the strings $\langle V_x^s\rangle$
vs $N_s$, which we plot in  
Fig.~3(a) for systems with the same parameters as in Fig.~1 at
$F_{D} = 3.0$, $4.0$, and $5.0$. 
The bottom curve, which
corresponds directly to the system in Figs.~1 and 2, 
shows that for $N_s = 1.0$, $\langle V_x^s \rangle=0.35$,
while for $N_s = 5$ and $6$, $\langle V_x^s \rangle \approx 0.15$. 
The string velocity increases rapidly with $N_s$ for $N_s < 5$ and
then increases much more slowly with $N_s$. The longer strings become
considerably more winding and experience additional frictional drag,
whereas the shorter strings move rigidly.
Due to the meandering of the longer strings, there
is a tendency for the driven particles at the back 
of the string to be ripped away from the string very quickly.
For curves with higher $F_{D}$, 
there is still a  
monotonic increase in the average velocity with $N_s$.
In order to show that this phenomenon is robust over a wide range 
of parameters, in Fig.~3(b) we
plot $\langle V_{x}^s\rangle$ vs $N_s$ for a system with $q_{D}/q = 1.33$ for 
$F_{D} = 0.5$, $0.75$, and $1.0$. 
For these parameters we again observe the string formation 
and an increase in $\langle V_{x}^s\rangle$ 
with $N_s$. The best functional fit we find 
is $\langle V_{x}^s\rangle \propto \ln(N_s)$; 
however, other forms can also be fit, such as 
two linear fits for the system with $q_{D}=1.33$.

In Fig.~4(a) we show the most probable number of particles 
$\langle N_s\rangle $ in a string for
the same system in Fig.~1 for varied
$F_{D}$. 
Here $\langle N_s\rangle$ 
goes through a maximum as a function of $F_{D}$. As $F_D$ decreases
below $F_D<0.2$,
the distortion created by the driven particles in the surrounding
background particles decreases, and therefore there is less 
attraction between the driven particles.  
At high $F_D$, the velocity fluctuations along the string 
become important, which tends to limit the string length. 
In Fig.~4(b) we show that $\langle N_s\rangle$ for 
a system with a fixed $F_{D} = 1.0$ also goes through a maximum as a
function of $q_{D}/q$.  For small $q_{D}/q$, the driven
particles produce very little distortion in the 
surrounding medium so there is 
little attraction between them. For
large $q_{D/q}$, the repulsion between the driven particles becomes 
more prominent and the average velocity 
drops, which again reduces the distortion field.
In Fig.~4(c) we plot $\langle N_s\rangle$ vs 
the particle density $n_c$ for a system with $q_{D} = 6.67$ and 
$F_{D} = 2.0$. At low $n_c$, all the particles are far apart and the 
distortion in the bath particles is reduced or absent. 
As $n_c$ increases,
the particle-particle interactions are enhanced and the string 
can grow longer. Presumably for 
very high $n_c$, $\langle N_s\rangle$ will decrease again 
as the average velocity will have to be reduced.
We are not able to access particle densities in this range. 

In Fig.~4(d) we plot $\langle V_x^s\rangle$  vs the 
system size $L$ for 
$F_D=1.5$ (lower curve), 3, 4.5, and 6.0 (upper curve).
In each case, 
$\langle V_x^s\rangle$ 
initially decreases with $L$ but reaches a saturation value by
$L\approx 40$.
The change in $\langle V_x^s\rangle$ at small $L$ occurs because, in  
small systems, the periodic boundary conditions allow driven colloids 
to interact strongly with their own distortion trails. 
For larger $F_{D}$, larger system sizes must be used to avoid this effect.   
In addition to the increase in $\langle V_x^s\rangle$ in the small systems, 
$\langle N_s\rangle$
also increases for the smallest systems. 
This emphasizes the fact 
that sufficiently large systems must be considered in order
to avoid finite size effects. 
We have used $L=48$ throughout this work,
which is large enough to avoid finite size
effects for the 
quantities we have measured.  

The attraction between the driven particles 
is mediated by the 
distortion in the surrounding bath of colloids.
The concept of nonequilibrium 
depletion forces, which are highly anisotropic and have both an
attractive and a repulsive portion,
was introduced recently for a   
case where two large colloidal particles have
an additional bath of smaller particles flowing past them
\cite{Likos}. 
The surrounding particle density is 
higher in front of the large particle 
and lower in back, which causes an anisotropic 
attraction force between the large colloids that 
aligns them with the direction of flow.
Although this
model was only studied for two particles, it seems to 
capture several of the features that we find in our simulations, including
the alignment of the strings in the direction of drive. 
This suggests that nonequilibrium depletion
force models could be applied for multiple large particles. 
The longer strings move faster since the front particle has 
additional particles pushing on it
from the back. 
Since the chains are aligned, the cross section with the 
background particles remains the
same as for a single driven particle; however, the string is 
coupled $N_s$ times more strongly to the external field. 
As longer strings become
increasingly meandering, the cross section with the 
background particles increases 
and limits the speed that long strings can attain. 

In conclusion, we have investigated a system of repulsively interacting 
Yukawa particles
which form a disordered background for a small fraction of Yukawa 
particles that 
couple to an external drive. In this system, despite 
the fact that all of the pair-wise interactions are
repulsive, there is an emergent effective attraction between the 
driven particles which results in the formation of
one-dimensional intermittent stringlike structures. 
The velocity of a string increases with the 
number of particles in the string. 
The attraction between the driven particles 
is mediated by the distortion of the 
surrounding particles and is likely  
a result of the recently proposed nonequilibrium depletion forces. 
Longer strings move faster than shorter strings due to
the additional pushing force from the back  particles in the string. 
This system represents a new class of collective transport 
where multiple interacting
colloids are driven through a deformable 
disordered background, which is distinct from the case where
the background disorder is quenched. 

We thank E. Ben-Naim for useful discussions.
This work was supported by the U.S. DoE under Contract No.
W-7405-ENG-36.

\end{document}